\documentclass[aps,prb,preprint]{revtex4}
\usepackage{amsmath,amsfonts,amssymb}
\usepackage{graphicx,subfigure,epsfig}

\usepackage{color}
\definecolor{highlight}{RGB}{220, 80, 80}

\usepackage{siunitx}

\begin{document}

\title{Real time phase imaging with an asymmetric transfer
function metasurface}

\author{Lukas Wesemann}
\affiliation{School of Physics, University of Melbourne, Victoria 3010, Australia}
\affiliation{ARC Centre of Excellence for Transformative Meta-Optical Systems, School of Physics, University of Melbourne, Victoria 3010, Australia}
\author{Jon Rickett}
\affiliation{School of Physics, University of Melbourne, Victoria 3010, Australia}
\author{Timothy J. Davis}
\affiliation{School of Physics, University of Melbourne, Victoria 3010, Australia}
\author{Ann Roberts}
\email[Corresponding author:]{ann.roberts@unimelb.edu.au}
\affiliation{ARC Centre of Excellence for Transformative Meta-Optical Systems, School of Physics, University of Melbourne, Victoria 3010, Australia}
\affiliation{School of Physics, University of Melbourne, Victoria 3010, Australia}

\begin{abstract}
The conversion of phase variations in an optical wavefield into intensity information is of fundamental importance for optical imaging technology including microscopy of biological cells. While conventional approaches to phase-imaging commonly rely on bulky optical components or computational post processing, meta-optical devices have recently demonstrated all-optical, ultracompact image processing methods. Here we describe a metasurface that exploits photonic spin-orbit coupling to create an \emph{asymmetric optical transfer function} for real time phase-imaging. The effect of the asymmetry on transmission through the device is demonstrated experimentally with the generation of high contrast pseudo-3D intensity images of phase variations in an optical wavefield without the need for post-processing. This non-interferometric method has potential applications in biological live cell imaging and real-time wavefront sensing.
\end{abstract}

\maketitle
While the intensity of an optical field is readily measured with conventional camera technology, some applications require information beyond the intensity, such as polarization and phase. Phase-imaging is fundamental to imaging live, unstained biological cells and permits the investigation of cell dynamics and the detection of disease \cite{Park2018}. Conventional all-optical, phase-imaging methods, including differential interference contrast (DIC) \cite{arnison2004linear}, Zernike phase-contrast microscopy \cite{zernike1942phase} and other interference-based approaches, require bulk-optical components and light progragation over macroscopic distances, rendering them unsuitable for integration into compact optical devices. Other non-interferometric methods, such as those based on the transport-of-intensity equation (TIE)\cite{ampem2008index} and ptyography, \cite{marrison2013ptychography} require additional computational post-processing. Recently meta-optical devices ranging from photonic crystals and thin-film structures through to optical metasurfaces have gained significant scientific attention as platforms for all-optical computation including image processing \cite{silva2014performing,zhu2017plasmonic,Dong2018,guo2018photonic,bykov2018first,Cordaro2019,Zhu2019,He2020,Zhou2020m,zhou2020flat}. There has been a strong focus on the application of such systems to all-optical enhancement of sharp edges in intensity and phase images \cite{zhu2017plasmonic,Wesemann2019,zhou2020flat}. The visualization of gradual phase-changes in a wavefield, on the other hand, has attracted less attention despite its importance for sensing spatial wavefront variations and the visualization of other features in transparent objects beyond their edges \cite{Vohnsen2015,zhu2020optical}. The application of a meta-optical device to phase contrast imaging of biological cells has been recently reported \cite{wesemann2021nanophotonics}. To create pseudo-3D contrast, this device required a small tilt ($\approx 3^\circ$) with respect to the optical axis of the imaging system to break the symmetry in its optical transfer function, introducing additional complexity. Here we experimentally demonstrate an optical metasurface that exploits the phenomenon known as photonic spin-orbit coupling \cite{Bliokh2015} to generate the required asymmetry about normal incidence, i.e. without tilting the device, and present the first demonstration of its application to ultra-compact, all-optical phase-imaging. More generally, the introduction  of asymmetry in the optical transfer function of metasurface devices is important to enable more complex spatial filtering for feature detection and visualization enhancement.

It is well known that filtering spatial frequencies in a wavefield can lead to phase contrast. The ability of an optical system to manipulate the Fourier content of a wavefield is described by its optical transfer function (OTF).
\begin{figure}[ht]
    \centering
    \includegraphics[width=0.9\linewidth]{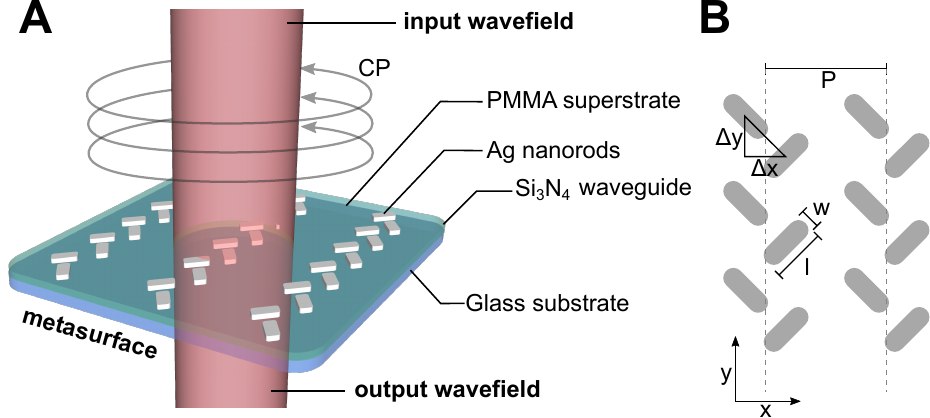}
    \caption{Phase imaging using spin-orbit coupling enabled by plasmonic metasurface. (a) Concept and schematic of structure (b) Geometric parameters of array of Ag nanorods.}
    \label{fig:Fig1}
\end{figure}
We consider a scalar, monochromatic, spatially coherent wavefield that has a pure phase-modulation $E(t,x,y,z=0)=E_0e^{i\phi(x,y)}e^{-i\omega t}$ where $\phi(x,y)$ describes the phase variation in the $z=0$ plane, $E_0$ is a constant and $\omega$ is the angular frequency of the wave. Since $I\propto|E|^2=|E_0|^2$, the intensity of this field carries no information about the phase distribution. The Fourier content of the field, however, can be manipulated to enable visualization of the phase \cite{zernike1942phase}. We elaborate on this further in supplement SI1.

Common phase imaging techniques such as differential interference contrast (DIC) distinguish positive and negative phase gradients in a wavefield, thereby generating characteristic pseudo-3D intensity images \cite{Wesemann2019}. In the Fourier domain, the key to achieving this is the ability of the phase-imaging system to distinguish plane waves propagating in directions at angles on either side of the propagation axis. This is expressed as an asymmetry in the OTF of the system \cite{PhysRevLett.123.013901}. In previous research into image processing such an asymmetry in the transfer function of a meta-optical component has been generated  by tilting the device with respect to the optical axis \cite{zhu2017plasmonic,bykov2018first,wesemann2021nanophotonics}. Optical spin-orbit effects result from a coupling between the spin angular momentum of a light beam and coordinate frame rotations associated with an optical medium giving to phase effects \cite{Bliokh2015} and it is this approach we use here to produce the required asymmetry in the OTF. 

Here we consider a metasurface that consists of an array of Ag nanorods fabricated on a $\SI{150}{\nm}$ thick Si$_3$N$_4$ waveguide sandwiched between a glass substrate and a PMMA superstrate (Fig. \ref{fig:Fig1}a). The nanorods of length $l$ and width $w$ are arranged in columns with a period $P$ as shown in Fig. \ref{fig:Fig1}b. Each column consists of two sublattices spaced by $\Delta x$ in which the nanorods are oriented at $45^\circ$ and $135^\circ$ to the $x$-axis. Fradkin et al. \cite{fradkin2020nanoparticle} showed numerically that this geometry exhibits a transmission asymmetry about normal incidence when illuminated with circularly polarized light. The change in orientation of the nanorods with position appears as a coordinate transformation with respect to the long axes of the rods, in this case a rotation, that varies across the glass surface. Circularly polarised light has an intrinsic spin angular momentum related to the rotation of its electric field vector. Interaction of this spin with the different orientations of the nanorods leads to spin-orbit coupling that generates phase shifts creating an asymmetry in the response of the metasurface \cite{fradkin2020nanoparticle}. In particular, the incident light excites plasmonic dipole resonances that oscillate with a phase difference of $\Delta \phi_1=\pi/2$ due to the phase difference between the $p$- and $s$-polarized components of the incident CP light. Each of the sublattices acts as a grating coupler for the underlying waveguide layer transferring the phase shift of the rod resonances onto the propagating fields in the waveguide. For a separation between the sublattices of $\Delta x = P/4$, an additional phase shift of $\Delta \phi_2=\pi/2$ is generated between the waveguide modes excited via each sublattice. In this case each sublattice excites guided modes that interfere constructively in one direction, where the phase shifts compensate such that $\Delta \phi_1+\Delta \phi_2=0$, and destructively in the opposite direction, where $\Delta \phi_1+\Delta \phi_2=\pi$. This unidirectional mode propagation creates a process referred to as `directional coupling' \cite{lin2013polarization}. For oblique incidence with $\theta_{\mathrm{inc}} \neq 0$ the interference condition is offset, enabling an asymmetric angular response of the device. This interference generates the polarization-dependent asymmetric optical transfer function that underpins the phase-imaging method proposed here.

\begin{figure}[ht]
    \centering
    \includegraphics[width=0.9\linewidth]{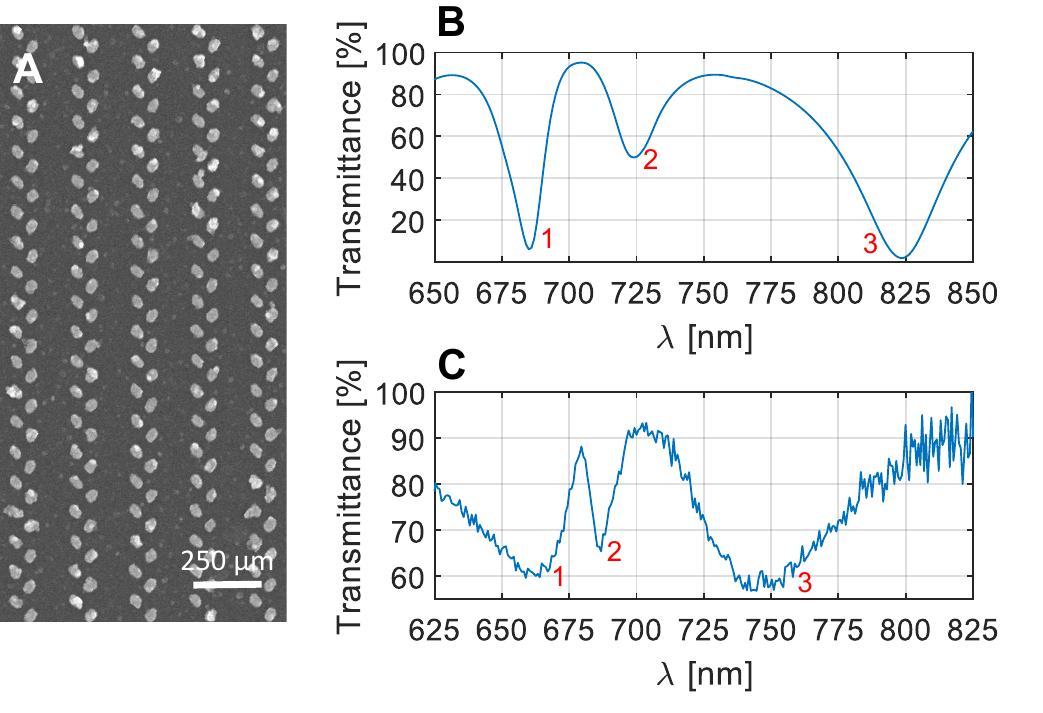}
    \caption{(a) SEM image of metasurface. (b) Measured and (c) calculated normal incidence transmission spectra for circularly polarized light with (c) obtained using geometric parameters from SEM analysis.}
    \label{fig:Fig2}
\end{figure}

The metasurface investigated here was designed for operation in the red and near-infrared spectral regions and fabricated through an electron beam lithography process that is described in the supplementary material SI2. A scanning electron microscopy (SEM) image of the surface of the device is shown in Fig. \ref{fig:Fig2}a. The coupling mechanism of the device is highly sensitive to the geometric parameters of the nanorods. For this reason we obtained averaged values from the SEM images for inclusion in simulations. The resulting parameters for the rod dimensions are $l=(100\pm 10)\SI{}{\nm}$ and $w=(64\pm 9)\SI{}{\nm}$, with the grating periodicity in the $x-$ and $y-$directions $P_x=\SI{420}{\nm}$, $P_y=\SI{210}{\nm}$ and the spacing of the rods $\Delta y=\Delta x=P_x/4=\SI{105}{\nm}$. 

The optical response of the metasurface is modelled using the finite element method (FEM), as described in supplement SI3. Fig. \ref{fig:Fig2}b shows the calculated transmission spectrum of the device.
\begin{figure*}[ht]
    \centering
    \includegraphics[width=0.90\linewidth]{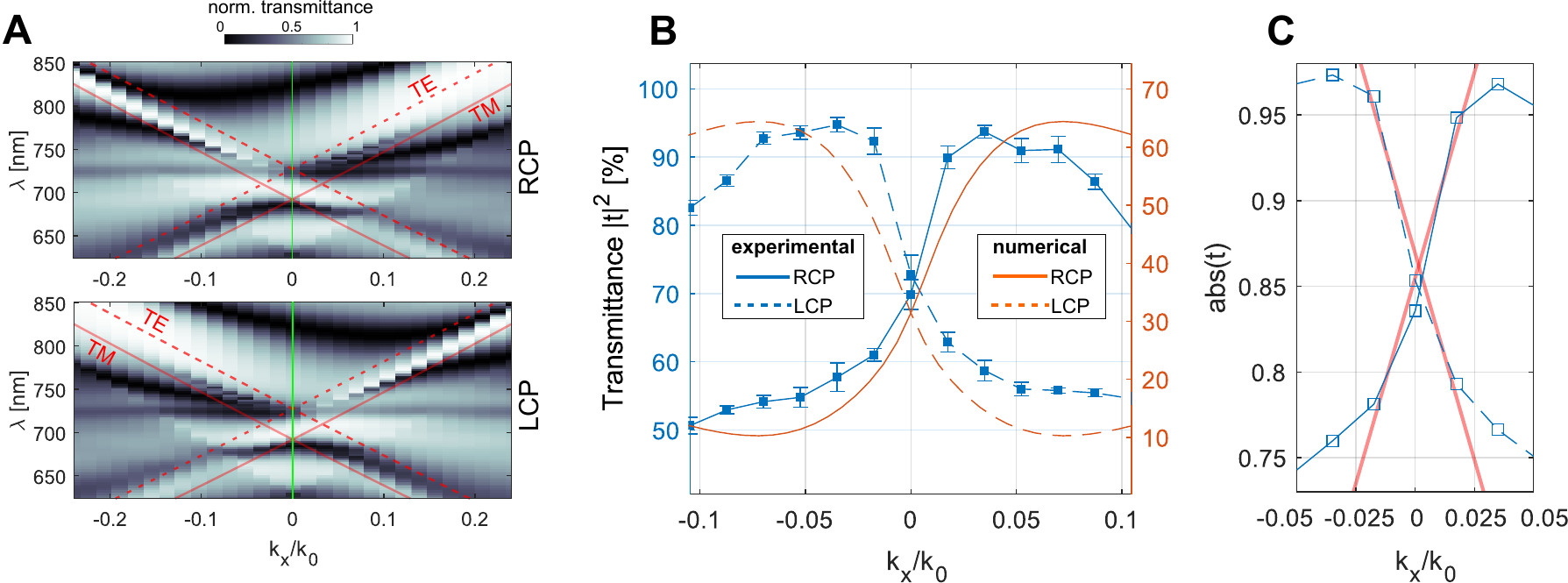}
    \caption{Angular asymmetry for transmission of circularly polarized light through the metasurface. (a) Calculated transmittance through the metasurface as a function of wavelength $\lambda$ and spatial frequency $k_x/k_0$ for RCP and LCP illumination (b) Experimental (blue) and calculated (orange) $|H(k_x/k_0)|^2$ at excitation of resonance feature '2'. (c) Measured MTF $|H(k_x/k_0)|$ (blue) at excitation of resonance feature '2' with linear fit (red) over the interval [$-0.025<k_x/k_0<0.025$].}
    \label{fig:Fig3}
\end{figure*}
In this configuration the metasurface exhibits three spectral features, with central wavelengths of $\lambda_1 = \SI{685}{\nm}$, $\lambda_2 = \SI{724}{\nm}$ and $\lambda_3 = \SI{824}{\nm}$, that suppress transmission at normal incidence for circularly polarized (CP) light  (Fig.\ref{fig:Fig2}b). These features arise from the strong coupling between the TE and TM waveguide modes and the plasmonic electric dipole resonances of the silver nanorods. Experimental measurement of the transmission spectrum through this metasurface shown in Fig.\ref{fig:Fig2}c, reproduces these spectral features. We attribute the shift in the experimentally measured resonances compared to simulations at $\lambda_1 = \SI{662}{\nm}$, $\lambda_2 = \SI{687}{\nm}$ and $\lambda_3 = \SI{745}{\nm}$ to differences in the dielectric properties and details of the geometry of the fabricated Si$_3$N$_4$ layer and Ag nanorods as discussed previously. Furthermore, inevitable fabrication inaccuracies arising from the electron beam lithography process are apparent in Fig.\ref{fig:Fig2}a. Both of these effects alter the plasmonic resonances of the rods as well as the dispersion of the underlying waveguide layer leading to shifts and broadening of the metasurface resonances.

In Fig.\ref{fig:Fig3}a the simulated transmittance through the device is shown as a function of normalized transverse spatial frequency $k_x/k_0$ and free space wavelength $\lambda$ with $k_x = k_0\sin(\theta_{\mathrm{inc}})$ and $k_0=2\pi/\lambda$. Here $\theta_{\mathrm{inc}}$ is the incidence angle of the input beam relative to the surface normal in air. An asymmetric response between positive and negative spatial frequencies is apparent that reverses with a change of the helicity of the circularly polarized incident light.  The red lines overlaid on Fig.\ref{fig:Fig3}a indicate the dispersion of the fundamental TE and TM modes of the slab wavguide, with any variation due to the presence of the plasmonic grating not taken into account. It is apparent that the resonance features of the metasurface are associated with the excitation of the TE and TM modes. Differences between the resonances identified in the FEM calculations and the simple slab waveguide dispersion relations can be attributed to a significant change in the waveguide dispersion due to the presence of the metallic grating and strong coupling between the resonances of the plasmonic nanorods and the waveguide modes. To maximize the sensitivity of the device to spatial frequency we selected operation at the wavelength of the spectral feature marked '2' in Figs. \ref{fig:Fig2}b,c.

The capacity of the device for asymmetric spatial frequency filtering was experimentally demonstrated by recording transmission spectra as a function of angle of incidence (Fig. \ref{fig:Fig3}a). A  collimated incident beam from a tungsten halogen lamp is circularly polarized with a linear polarizer and a quarter-wave plate with their respective optical axes oriented at an angle of $\pm 45^\circ$ to obtain LCP or RCP illumination respectively. The transmitted light is collected with a microscope objective and fed into a spectrometer via fiber coupling. Further information about the experimental setup can be found in the supplementary material (SI4). The transmission spectra are normalized to the reference transmittance through an unpatterned region of the sample. In Fig. \ref{fig:Fig3}b the measured and calculated transmittance is shown for both LCP and RCP illumination at the wavelength of spectral feature '2' - 687 nm. It is apparent that the device provides asymmetric filtering of the Fourier content of an incident field at this wavelength along the $k_x$-direction. This filter function is essentially one-dimensional with further discussion provided in the supplement (SI1). As discussed, fabrication imperfections lead to the experimentally measured transmittance exhibiting a lower angular filtering contrast than obtained from the FEM result translating to weaker phase-contrast in imaging applications.

To avoid artifacts in the intensity image, the modulation transfer function (MTF) of the system $|\mathcal{H}(k_x/k_0)|=|t|$ requires a near linear dependence on spatial frequency, accompanied by a negligible variation in the phase transfer function $\arg{(\mathcal{H}(k_x/k_0))}$, near the origin. For the metasurface, the experimentally measured MTF shown in Fig. \ref{fig:Fig3}c is near-linear over the approximate operational interval [$-0.025<k_x/k_0<0.025$], which is sufficient for phase-imaging of micrometer sized phase objects as demonstrated below. The phase-imaging performance of the metasurface is given by both its numerical aperture, NA$ = 0.025$, as well as the achievable contrast $\approx 40\%$ of $|H(k_x/k_0)|^2$ across its operational interval It is important to note that the NA of the metasurface refers to the interval of spatial frequencies where the object Fourier transform experiences a linear modification, while the resolution of the imaging system, in which the metasurface is used, will ultimately be given by the NA of the complete optical system.\\
\begin{figure*}[ht]
    \centering
    \includegraphics[width=1\linewidth]{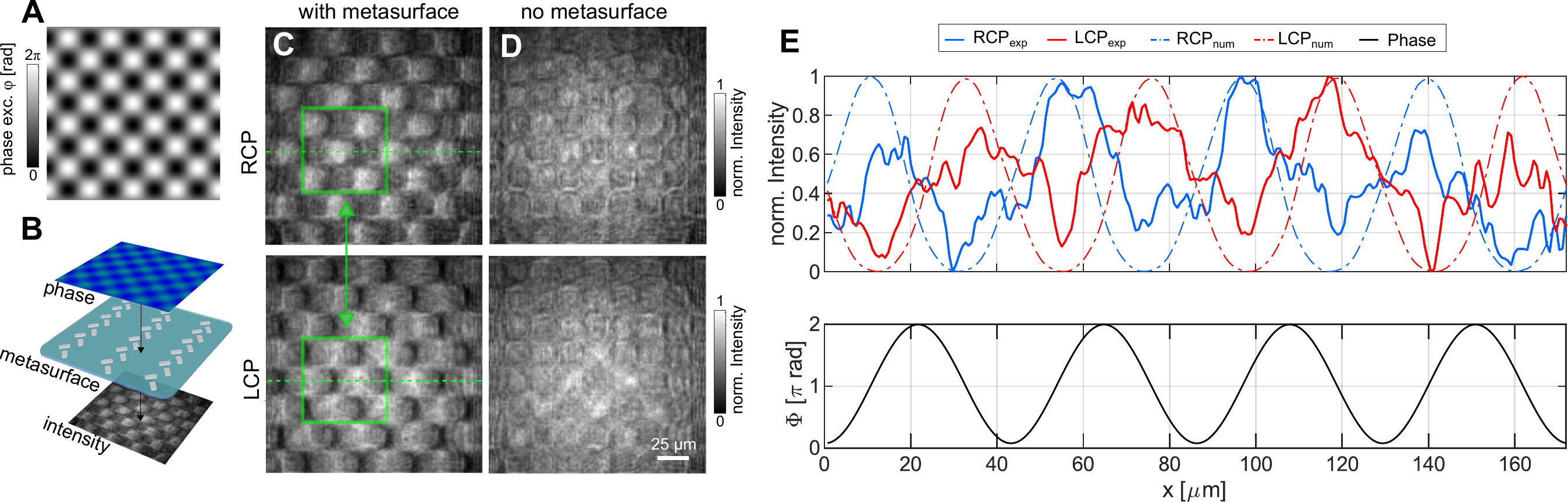}
    \caption{Phase imaging of eggcrate pattern using the photonic spin-orbit coupling metasurface at resonance feature '2'. (a) Phase excursion used as phase-object on spatial light modulator (SLM). (b) Concept of phase-to-intensity conversion via transmission through the metasurface (c) Obtained intensity image for RCP and LCP light producing pseudo 3D image of incident phase-object. Green insets highlight the 'flipping' of the image when the polarization state is changed from RCP to LCP (d) Reference images obtained for transmission of the phase-object through an unpatterned region of the sample. (e) Lineplot through the dashed green line of the experimental result in (c) (solid) and comparison with calculated transmission (dashed).}
    \label{fig:Fig4}
\end{figure*}
Using this device we demonstrate phase-contrast imaging by transmitting a circularly polarized wavefield with a superimposed phase modulation through the metasurface and recording the transmitted intensity image (Fig. \ref{fig:Fig4}b). The phase modulation is generated with a computer controlled spatial light modulator (SLM) with full experimental details provided in the supplementary material (SI5). The phase pattern used in the experiment takes the form of a eggcrate pattern with a normalized spatial frequency of 0.017 superimposed on a uniform background (Fig. \ref{fig:Fig4}a). The spatial frequency content of the projected phase distribution hence lies in the linear region of the transfer function. The intensity images obtained upon transmission through the metasurface for incident RCP and LCP light show clear pseudo-3D representations of the phase profile for both polarization states (Fig. \ref{fig:Fig4}c,d). These phase images highlight regions of increasing (RCP) and decreasing (LCP) edges of the phase modulation. The edge of the pattern associated with an intensity maximum can be switched via the handedness of the circular polarisation. This effect provides an additional degree of freedom to adjust the phase-contrast. A change from RCP to LCP illumination, therefore, leads to a `contrast reversal' in the pseudo-3D images as indicated in the green inset. Reference images of the wavefield transmitted through an unpatterned region of the substrate are shown on the right side in Fig. \ref{fig:Fig4}c,d. With the exception of minor intensity variations attributed to imperfections in the imaging system and the SLM, the phase-modulation is invisible in these reference images. In Fig. \ref{fig:Fig4}e lineplots along the green dashed lines in Fig. \ref{fig:Fig4}c,d are shown together with the numerically calculated intensity variation for filtering of the of the incident field with the calculated OTF of this metasurface.

Here we have demonstrated the first experimental application to phase contrast imaging of an optical metasurface with an asymmetric transfer function using normally incident light. We showed that the use of photonic spin-orbit coupling to create an asymmetric optical transfer function enables real-time phase-imaging of transmitted wavefields creating pseudo 3D intensity images using circularly polarized light. This new approach opens up intriguing prospects for biological phase imaging. Transparent objects, including live biological cells, placed directly on the metasurface would, in principle, generate intensity relief images similar to those obtained using differential interference microscopy (DIC)\cite{wesemann2021nanophotonics}. Similarly, this approach could be used  to visualize refractive index or thickness variations in other transparent media. The real-time, all-optical, phase-imaging properties of the metasurface could be exploited in time-critical applications such as high throughput flow cytometry \cite{picot2012flow} or wavefront sensing. Access to asymmetric optical transfer functions is also essential for a broader set of operations in all-optical computation, underpinning mathematical operations including first-order spatial differentiation. Such operations are necessary in high-throughput, real-time image processing situations such as autonomous vehicles, and their execution in the optical domain could alleviate speed and energy consumption limits of electronic signal processing. Finally, the integration of this metasurface approach with photodetectors could enable cost efficient and all-optical image processing for existing camera technology without the requirement for additional bulk-optical components \cite{panchenko2016plasmonic}.

\subsection*{Funding} The authors acknowledge funding through the Australian Research Council Discovery Projects Scheme (DP160100983) and the Center of Excellence Scheme (CE200100010). This work was performed in part at the Melbourne Centre for Nanofabrication (MCN) in the Victorian Node of the Australian National Fabrication Facility (ANFF).
\subsection*{Supplemental document}
See Supplement 1-5 for supporting content.


\end{document}


\setpagewiselinenumbers

\begin{center}
\section*{Real time phase imaging with an asymmetric transfer function metasurface}
\section*{Supplementary Information}
Lukas Wesemann$^{*,1,2}$, Jon Rickett$^1$, Timothy J. Davis$^1$, Ann Roberts$^{*,1,2}$\\
$^1$School of Physics, University of Melbourne, Victoria 3010, Australia\\ $^2$ARC Centre of Excellence for Transformative Meta-Optical Systems, School of Physics, University of Melbourne, Victoria 3010, Australia
\end{center}
\pagebreak

\subsection*{SI1. Spatial frequency filtering}

\subsubsection{Visualization of optical phase through spatial frequency filtering}
Filtering of the spatial frequency content of an optical field permits the visualization of the phase of a wavefield. To illustrate this we consider a scalar, monochromatic, spatially coherent wavefield with a pure phase-modulation $E(t,x,y,z=0)=E_0e^{i\phi(x,y)}e^{-\omega t}$. Here $\phi(x,y)$ describes the phase variation in the $z=0$ plane, $E_0$ is a constant and $\omega$ is the angular frequency of the wave.
The first spatial derivative of a wavefield can be computed through filtering of its Fourier content with a linear optical transfer function. The relationship between the input- and the output fields for a linear space-invariant optical system is given by $\tilde{E}_{\mathrm{out}}(k_x,k_y)=\mathcal{H}(k_x,k_y)\tilde{E}_{\mathrm{in}}(k_x,k_y)$, where $\tilde{E}_{\mathrm{in}}$ and $\tilde{E}_{\mathrm{out}}$ describe the spatial Fourier transforms of the input and output fields respectively and $\mathcal{H}$ the OTF of the system \cite{Goodman2005}. For a transfer function $\mathcal{H}(k_x,k_y) \propto k_x$, we then obtain a processed field $E_{\mathrm{out}} \propto \partial E_{\mathrm{in}}/\partial x$. When considering vector effects, the transfer function is replaced by a rank-2 tensor. For a nonlinear dependence of the transfer function on spatial frequency, which corresponds to higher order derivatives, a more complex relationship between phase modulation and resulting intensity is found. Although this permits a degree of visualization of the phase, it can result in artifacts in processed images \cite{Cordaro2019}.\\ \\
Of particular interest in phase imaging is it to obtain intensity images that allow to differentiate between positive and negative phase gradients. This is achieved for example by differential interference contrast microscopy (DIC) \cite{lang1982nomarski}, which produces typical pseudo 3D intensity images. In order to achieve this, an additional constant offset around which the optical transfer function is linear, is required. The spin-orbit coupling metasurface under consideration exhibits a near-linear magnitude transfer function around the origin with a constant offset making it suitable to produce pseudo 3D intensity images from incident phase images. Below we provide details on the optical transfer function of the metasurface.

\subsubsection{Spatial frequency filtering with spin-orbit coupling metasurface}
\begin{figure}[ht]
\centering
\includegraphics[width=\linewidth]{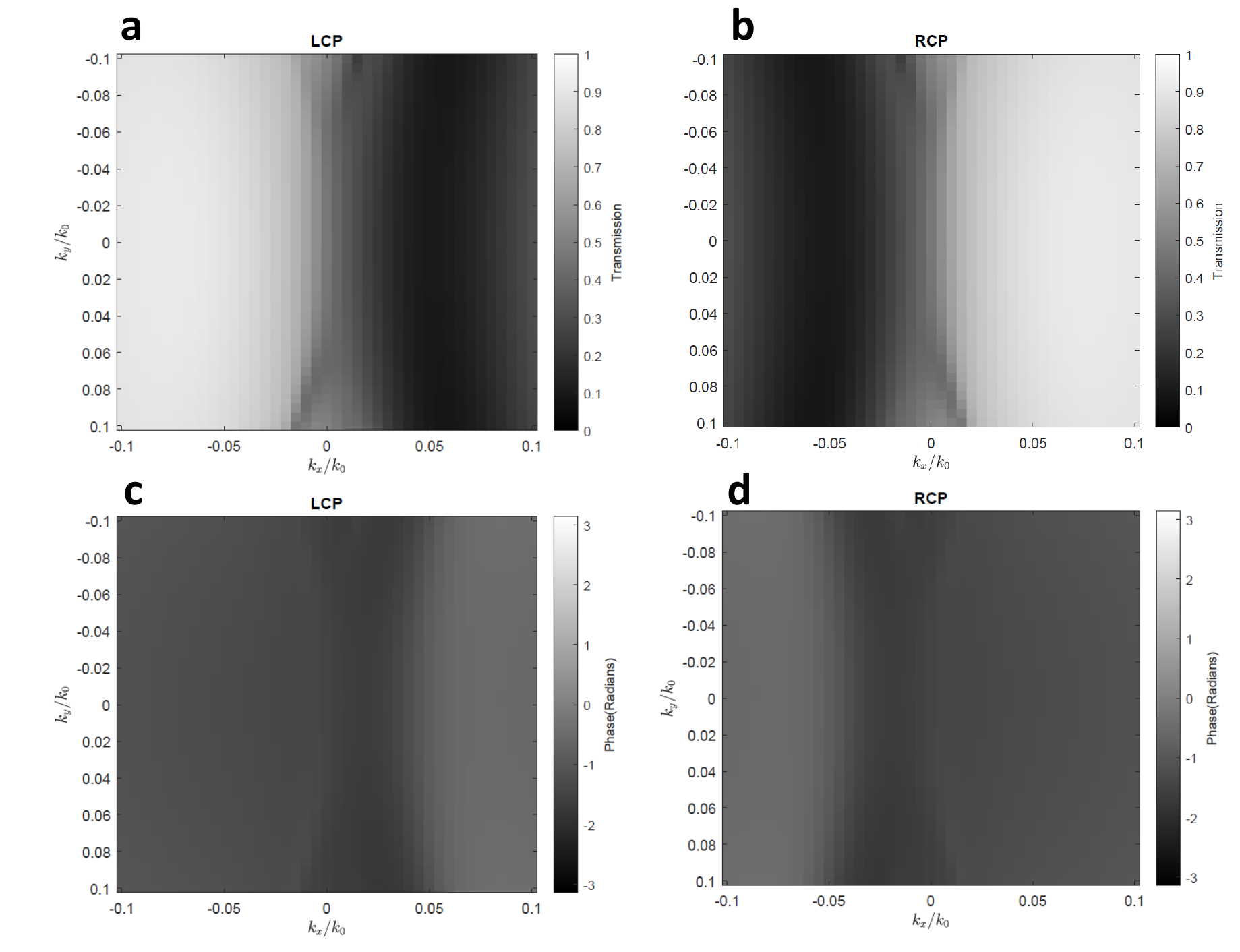}
\caption{Magnitude transfer function $|H(k_x/k_0, ky/k_0)|$ (a,b) and phase transfer function $\mathrm{arg}\big[H(k_x/k_0, k_y/k_0)\big]$ (c,d) of metasurface for RCP (first column) and LCP (second column) incident light.}
\label{fig:OTF}
\end{figure}

The metasurface under consideration enables asymmetric spatial frequency filtering in transmission using circularly polarized light. This capacity is characterized by numerically calculating the optical transfer function $H(k_x,k_y)$ of the device. Here we use the finite element method (FEM) as described further below in SI3. In figure \ref{fig:OTF} the magnitude transfer function $|H(k_x/k_0, ky/k_0)|$ (a,b) as well as the phase transfer function $\mathrm{arg}\big[H(k_x/k_0, k_y/k_0)\big]$ (c,d) are shown. It is apparent that the device essentially performs a one-dimensional filtering operation along the $k_x$ direction with minor dependence on $k_y$. In addition to this, the metasurface imposes a negligible phase-shift onto the transmitted field within the interval [$-0.025<k_x/k_0<0.025$] over which the metasurface exhibits a near-linear magnitude transfer function.

\subsection*{SI2. Metasurface Nanofabrication}
The metasurface was fabricated using a 
A $\SI{150}{\nm}$ thick layer of Si$_3$N$_4$ was deposited on a 4-inch glass wafer through plasma enhanced chemical vapour deposition (Oxford Instruments PLASMALAB 100 PECVD).  The metasurface pattern was defined by an electron beam lithography tool (Vistec EBPG 5000) in a single layer of polymethyl methacrylate resist (PMMA: 280nm A4, baked at $180^\circ$ C for $3$ min after deposition) that was spun onto the coated sample. The sample was developed in a 3:1 mixture of isopropanol: methylisobutyl. A 30 nm thick layer of Ag was subsequently deposited on the sample through physical vapor deposition on a $2$ nm adhesion layer of chromium. A $750$ nm thick layer of PMMA was subsequently spun on top of the nanorod array to ensure protection from degradation due to exposure to air. The PMMA was baked at $180^\circ$ C for $3$ min following the spincoating process.

\subsection*{SI3. Finite Element Calculations}
The metasurface is modeled using the finite element method (FEM) as implemented in COMSOL Multiphysics 5.5 with Wave Optics module. The model uses Floquet (periodic) boundary conditions in the transverse $x$- and $y$- directions and is terminated by port boundary conditions at the upper and lower boundary. Electromagnetic waves are launched into the model from the upper port with the $p$- and $s$- polarized components of the electric field $\pm \pi/2$ out of phase with respect to each other in order to model circularly polarized light. The optical transmission characteristics of the device are then obtained from the $S$-parameters at the output port. The optical constants for Ag and Si$_3$N$_4$ are obtained from \cite{Johnson1972} and \cite{luke2015broadband} respectively. The metasurface is assumed to be embedded in a homogeneous dielectric material of refractive index n=1.5, for both the substrate and the superstrate.

\subsection*{SI4. Optical Configuration for spectral measurements}
\begin{figure}[ht]
\centering
\includegraphics[width=11cm]{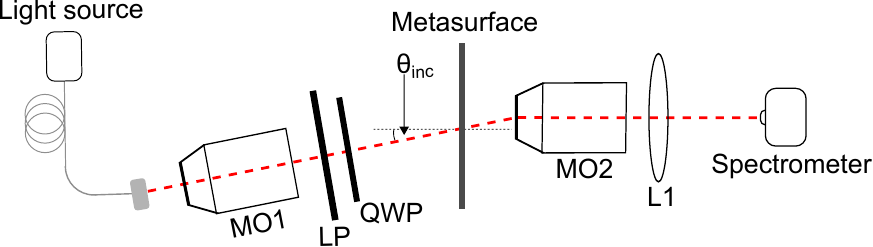}
\caption{Experimental setup used for the measurement of angle depended spectral transmission characteristics. Light from a tungsten halogen lamp is collimated and circularly polarized before it is transmitted through the metasurface at an adjustable angle of $\theta_{\mathrm{inc}}$. The transmitted light is guided to and analyzed by a spectrometer.}
\label{fig:setup_spectra}
\end{figure}
The optical configuration for the experimental spectral characterization of the metasurface is shown in Fig. \ref{fig:setup_spectra}. The setup was established on a Nikon Ti-80i inverted microscope using a tungsten halogen light source (Ocean Optics HL-2000 HP). Unpolarized light from the source collimated using a Nikon U PLAN 4x NA0.13 objective (MO1). The collimated beam is subsequently passed through a linear polarizer (Thorlabs LPVIS100) and a quarter-wave plate (Thorlabs AQWP05M-600) mounted on a rotation stage. This enabled adjusting the angle between the optical axes of the quarter-wave plate and the linear polarizer to $\pm 45^\circ$ thereby switching between right- and left circularly polarized input. The entire illumination unit (MO1, LP and QWP) was mounted on an $xyz$-stage with integrated rotation module to enable positioning of the input beam adjustment of the incident angle $\theta_{\mathrm{inc}}$ of the collimated beam onto the metasurface. The metasurface is placed on the integrated stage of the Nikon Ti-80i microscope. The transmitted light is collected by a Nikon LU PLAN LWD 50X 0.5NA objective (MO2). The light is then guided through the the microscope via the microscope's tube lens and subsequently fed into a single mode fiber (Thorlabs SM600) and guided to a spectrometer (Ocean Optics HR2000 + ES High-resolution spectrometer) where the spectral transmission data is recorded.

\subsection*{SI5. Optical Configuration for SLM Experiments}
\begin{figure}[htp!]
\centering
\includegraphics[width=11cm]{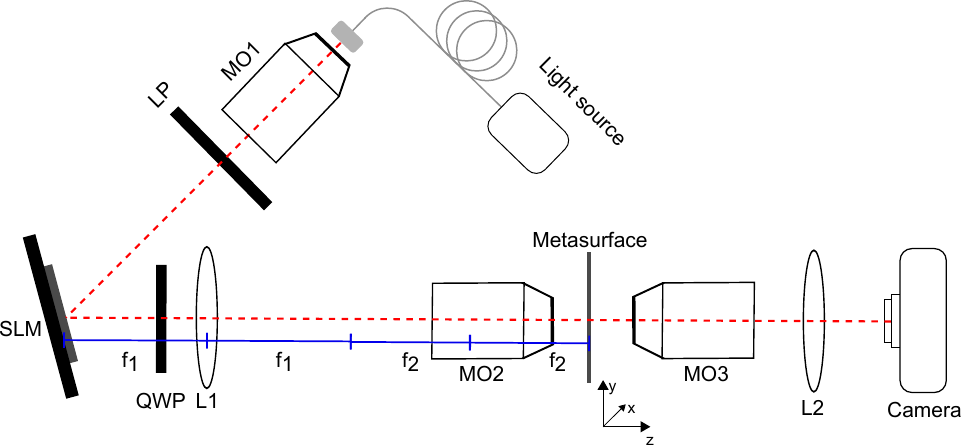}
\caption{Experimental setup used for the demonstration of phase-imaging via metasurface spin-orbit coupling. Phase-modulations are superimposed onto the wavefield using a spatial light modulation (SLM). The wavefield is circularly polarized, demagnified through a telescope system (L1,MO2) and projected onto the metasurface. The intensity converted image is recorded by a camera.}
\label{fig:setup_SLM}
\end{figure}
The phase-imaging experiments in this publication were carried out using the setup shown in Fig. \ref{fig:setup_SLM} that incorporates a spatial light modulator (Holoeye Pluto 2 VIS014) for the generation of phase-images. The spatial light modulator (SLM) consists of a $1920 \times 1080$ pixel liquid crystal on silicon display with pixel size of $\SI{8}{\mu m}$. Collimated light from a fiber coupled laser system (NKT Photonics SuperK Compact) with fiber-coupled tunable filter (NKTPhotonics SuperK Select Tunable Multi-line Filter) with a bandwidth of $3$ nm (FWHM) is linearly polarised along the operational direction of the SLM. The light reflected from the SLM is passed through a quarter-wave plate (Thorlabs AQWP05M-600) and subsequently demagnified through a telescope consisting of a $f=150$ mm  lens (Thorlabs-LA1433-A) (L1) and a micropscope objective (Nikon UPlanFl 20x 0.5NA) (MO2). The phase-image is projected onto the metasurface, which is mounted on an $xyz$-stage for precise adjustment. The transmitted image is subsequently captured using a microscope objective (Nikon LU Plan 50x 0.55NA) (MO3), and projected onto a camera (Thorlabs DCC1545M) through a $f=50$ mm lens (Thorlabs LA1131-A) (L2).
\pagebreak